\documentclass[a4paper,11pt]{article}
\usepackage{pos}

\usepackage{pifont}
\usepackage[skip=-1pt]{subcaption}
\usepackage{float}
\usepackage{siunitx}
\usepackage{wrapfig}
\usepackage[capitalize]{cleveref}
\usepackage[most]{tcolorbox}
\usepackage{enumitem}
\usepackage{tablefootnote}
\DeclareMathOperator{\tr}{tr}
\DeclareMathOperator{\diag}{diag}
\definecolor{buw_fac4_blue}{RGB}{0, 118, 175}

\title{Symmetry properties of staggered fermions with taste splitting mass term}

\author*[a]{Nuha Chreim}
\author[a]{Christian Hoelbling}

\affiliation[a]{Department of Physics, University of Wuppertal, Gaußstraße 20, D-42119 Wuppertal, Germany}

\emailAdd{nuha.chreim@uni-wuppertal.de}
\emailAdd{hch@uni-wuppertal.de}

\abstract{We present analytical and numerical results on symmetry properties of staggered fermions with taste splitting mass terms. As staggered species split differently for different types of taste splitting masses, various lattice symmetry subgroups and symmetry properties emerge. From the breaking of rotational symmetry, gluonic counterterms emerge. Preliminary numerical results are given for lattice sizes up to $8^4$.}

\FullConference{%
  The 41st International Symposium on Lattice Field Theory\\
  28th July - 3rd August, 2024\\
  Liverpool, England
}

\begin{document}
\maketitle

\section{Introduction}
The formulation of staggered fermions\cite{Susskind,Banks} as one of the first fermions on the lattice has advanced the simulation of fermion action early on and is still one of the widely used lattice fermion formulation to this day \cite{Fabi1,Boccaletti,Lahert,Verplanken,Fabi2}. Not only is the size of the naive fermion operator reduced by a factor of 4 while preserving a proxy of chiral symmetry; the distinct structure allows for many interesting formulations such as reduced staggered fermions \cite{Sharatchandra} or symmetric mass generation on the lattice \cite{Catterall,Butt}. In this work, we revisit the idea of including a taste dependent mass term, which decouples the doublers from the physical fermion in a similar manner as the Wilson term does for naive fermions \cite{Golterman, Adams, Hoelbling,DeForcrand,Duerr,Misumi,Zielinski}. We first explore the symmetry properties of staggered fermions with different mass terms, with an emphasis on rotational symmetry breaking, which causes the appearance of glounic counterterms \cite{Sharpe}. We provide preliminary numerical results in pure gauge theory. \\
Starting from the naive fermion action with the full fermion fields $\psi$, the gamma matrices $\gamma_{\mu}$ and a mass $m$
\begin{align}
    S_n = a^4 \sum_n \Bar{\psi}(n)\left(\gamma_{\mu}\frac{1}{2a}\left[\psi(n + \hat{\mu}) - \psi(n-\hat{\mu})\right]  + m\psi(n)\right)
\end{align}
and applying the so called staggered transformation of the fields
\begin{align}
    \psi(n) \to \Gamma(n) \chi(n),  \quad \Bar{\psi}(n) \to\Bar{\chi}(n) \Gamma^{\dagger}(n)
\end{align}
where $\Gamma(n) = \prod_{\mu}\gamma_{\mu}^{n_{\mu}}$, results in the staggered action
\begin{align}
    S_{st} =a^4\sum_n \Bar{\chi}(n)\left(\eta_{\mu}\frac{1}{2a}\left[\chi(n + \hat{\mu}) - \chi(n-\hat{\mu})\right]  + m\chi(n))\right)
\end{align}
with spinless fields $\chi$ and the staggered phase $\eta(n)=(-)^{\sum_{\nu < \mu}n_{\nu}}$. In leading order, the resulting staggered structure is encoded in the spin-taste basis \cite{Daniel}
\begin{align}
    (\gamma_{S}\otimes \xi_{F})_{xy}=\frac{1}{2^{D/2}}\tr(\Gamma^{\dagger}(x)\gamma_{S}\Gamma(y)\gamma^{\dagger}_{F})
\end{align}
where $\xi_{F}=\gamma_{F}^{T}$ acts on flavor indices. This basis suggests a splitting of the tastes into two pairs each by introducing spin singlet operators
\begin{alignat}{3}
    &\qquad 1\otimes \sigma_{\mu\nu} &&\longleftrightarrow M_{\mu\nu} =i\epsilon_{\mu\nu}\eta_{\mu}\eta_{\nu}C_{\mu}C_{\nu}\\
    &\qquad 1\otimes\xi_5 &&\longleftrightarrow M_A=\epsilon\eta_5C. 
\end{alignat}
where $\sigma_{\mu\nu}=i\xi_{\mu}\xi_{\nu}$. The 2-hop operator $M_{\mu\nu}$ involves an antisymmetric tensor $\epsilon_{\mu\nu}=(-)^{x_{\mu}+x_{\nu}}$ and hopping operators $C_{\mu}=\frac{1}{2}\left(U_{\mu}\delta_{x+\hat{\mu},y}+U_{\mu}^{\dagger}\delta_{x-\hat{\mu},y}\right)$. A single taste operator can be constructed \cite{Hoelbling} by a combination of  2-hop terms that hop in all four directions $M_{H}=M_{\mu\nu}+M_{\rho\sigma}$ with $\mu, \nu,\rho, \sigma$ permutations of $\{1,2,3,4\}$.  In the Adams term $M_{A}$ \cite{Adams},  $\epsilon=(-)^{\sum_{\mu}x_{\mu}}$ is the generator of the remnant chiral symmetry, $\eta_{5}=\eta_{1}\eta_{2}\eta_{3}\eta_{4}$ and $C=\left(C_{1}C_{2}C_{3}C_{4}\right)_{sym}$ is a symmetrized sum over all permutations of all hopping operators.

\section{Symmetry properties}
\begin{figure}[H]
            \centering
            \begin{subfigure}{0.5\textwidth}
                    \centering
                    \includegraphics[trim={5.5cm 0.8cm 3.5cm 0},clip,width=0.65\textwidth]{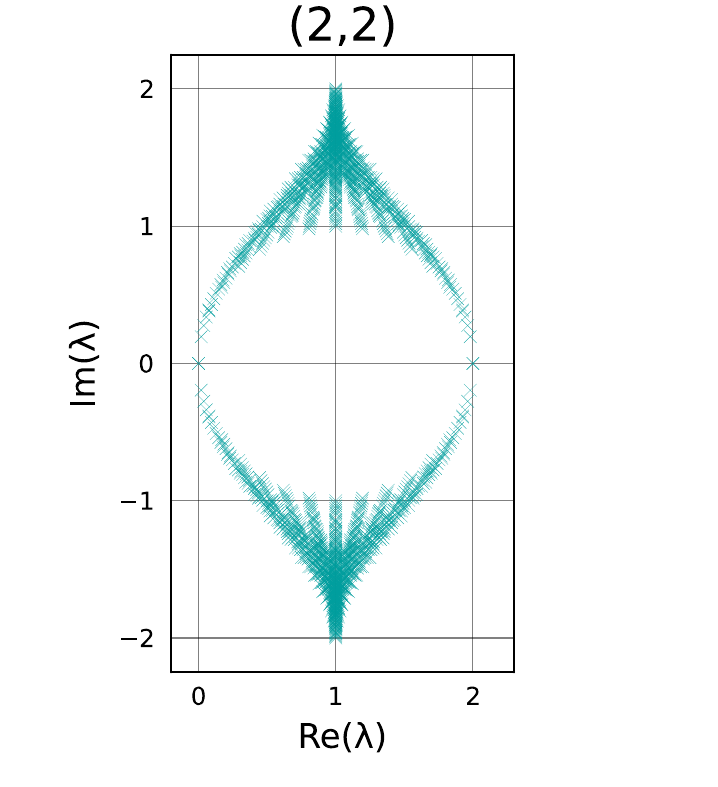}
                    \subcaption*{$M_{A}\sim \diag(-1,-1,1,1)$\\ \vspace*{0.2cm} 
                    $\{C_0, S_{\mu}I_S,R_{\mu\nu}\}$}
                    \label{fig:2-flavor_adams}
            \end{subfigure}
            \hspace*{-3cm}
            \begin{subfigure}{0.5\textwidth}
                \centering
                    \includegraphics[trim={5.5cm 0.8cm 3.5cm 0},clip,width=0.65\textwidth]{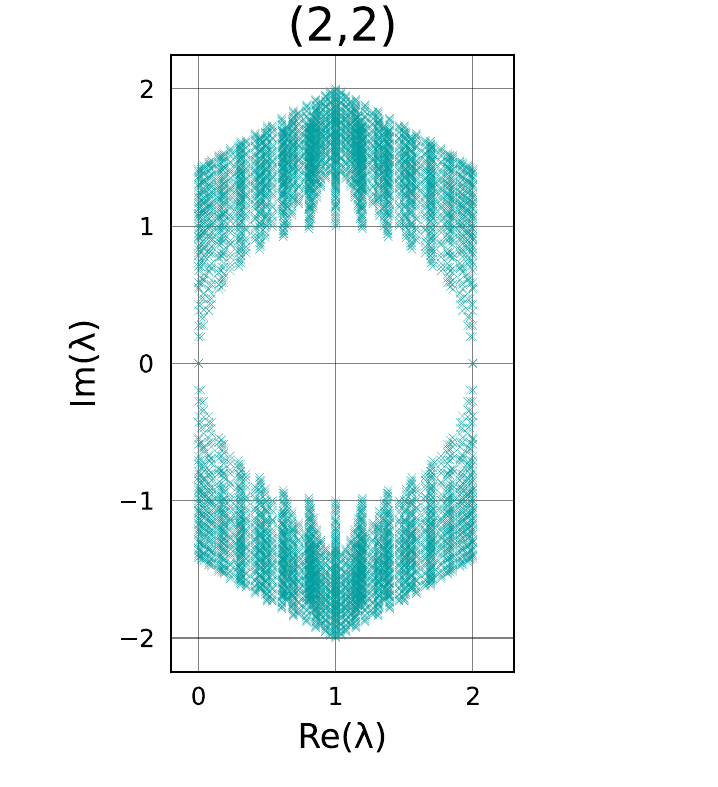}
                    \subcaption*{$M_{12}\sim \diag(-1,1,-1,1)$\\ \vspace*{0.2cm} 
                    $\{R^2_{13}C_0, R^2_{24}C_0,S_{\mu}I_S,R_{12},R_{34}\}$}
                    \label{fig:2-flavor_12}
            \end{subfigure}
            \caption{2-flavored mass terms and their symmetries by a 2+2 splitting.}
            \label{fig:2+2}
            \end{figure}

\begin{figure}[H]
            \captionsetup[subfigure]{justification=centering}
            \centering
            \begin{subfigure}{.5\textwidth}
                \centering
                \includegraphics[trim={3.5cm 0.8cm 2.5cm 0},clip,width=0.85\textwidth]{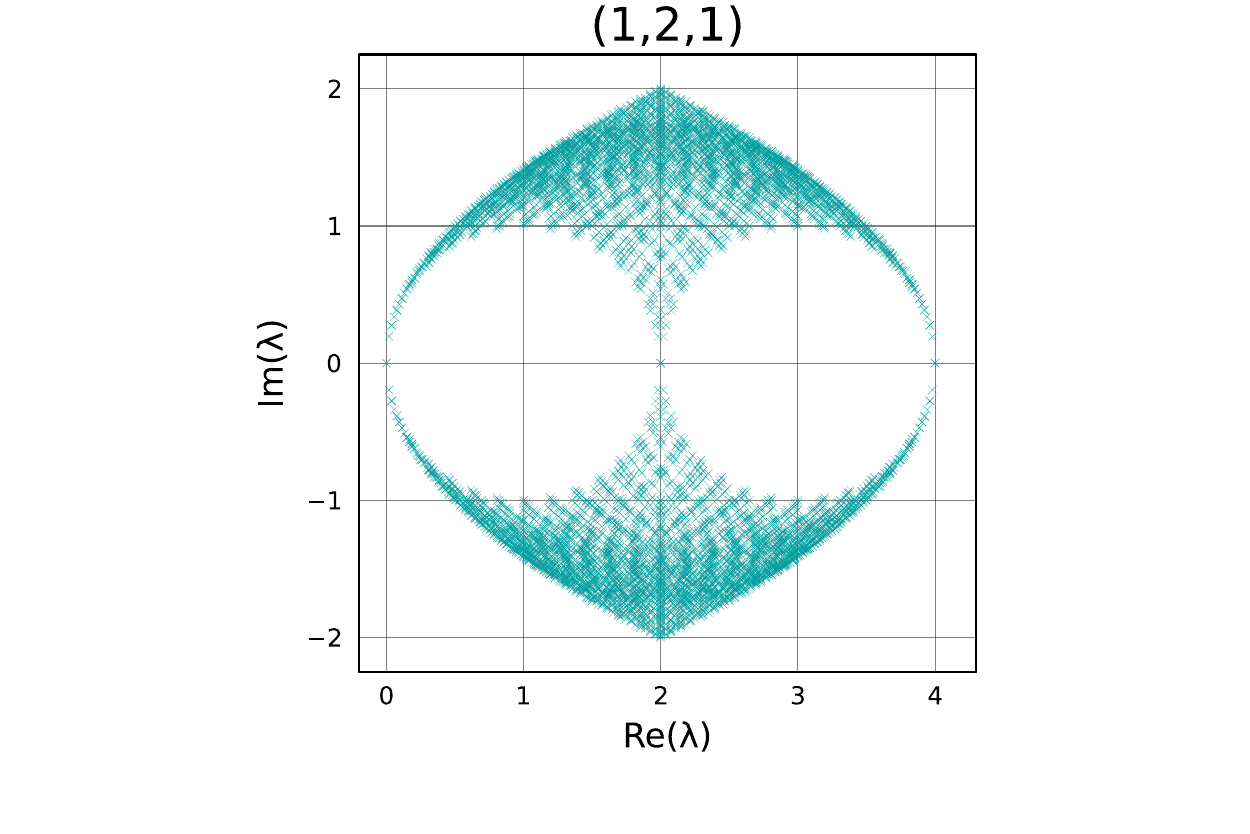}
                \subcaption*{ $M_{12}+M_{34}\sim\diag(0,0,-2,2)$ \\ \vspace*{0.2cm} 
                $\{R^2_{13}C_0, R^2_{24}C_0, S_{\mu}I_S,R_{\mu\nu}R_{\rho\sigma}\}$}
            \end{subfigure}%
            \hspace{-1cm}
            \begin{subfigure}{.5\textwidth}
                \centering
                \includegraphics[trim={3.5cm 0.8cm 2.5cm 0},clip,width=0.85\textwidth]{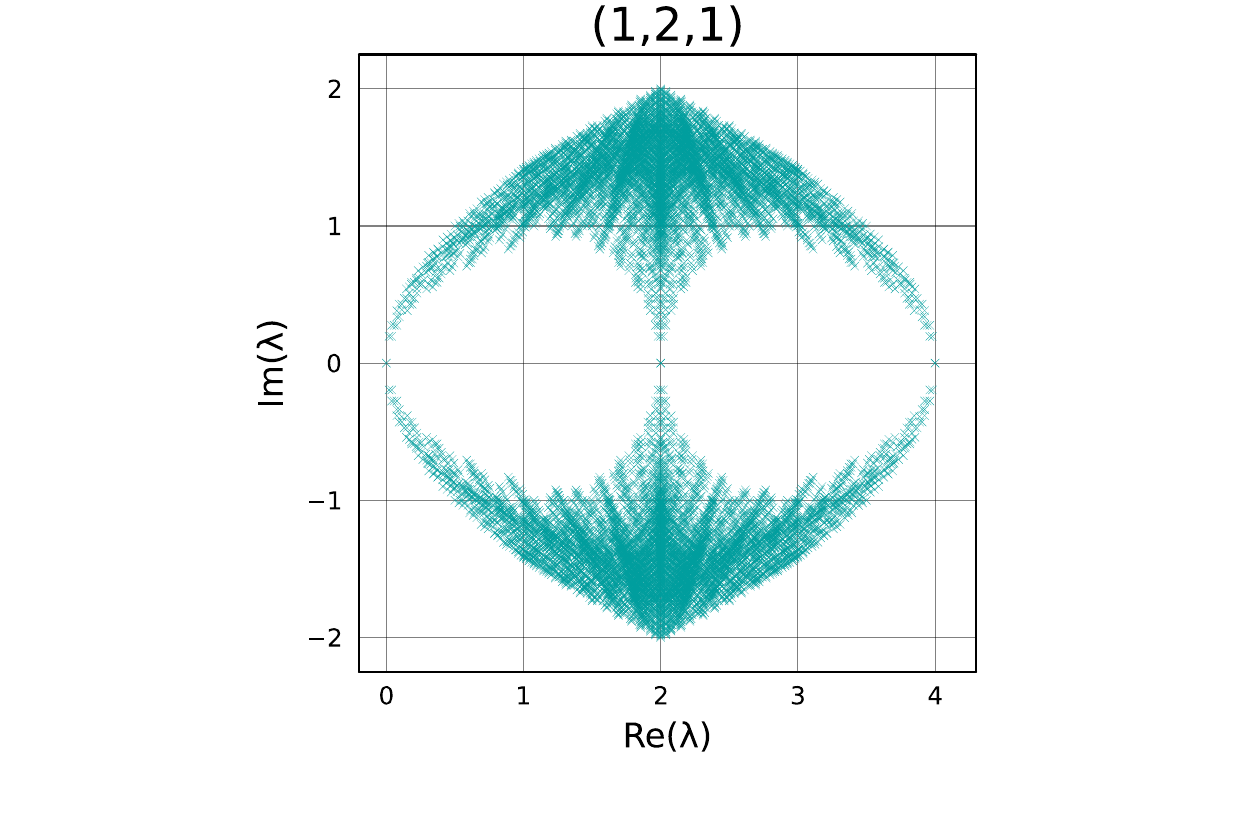}
                \subcaption*{$M_{A}+M_{12}\sim\diag(-2,0,0,2)$\\ \vspace*{0.2cm} 
                $\{R^2_{13}C_0, R^2_{24}C_0,S_{\mu}I_S,R_{12},R_{34}\}$.}
            \end{subfigure}%
            \par\bigskip
            \begin{subfigure}{.5\textwidth}
                \centering
                \includegraphics[trim={3cm 0.8cm 2cm 0},clip,width=0.85\textwidth]{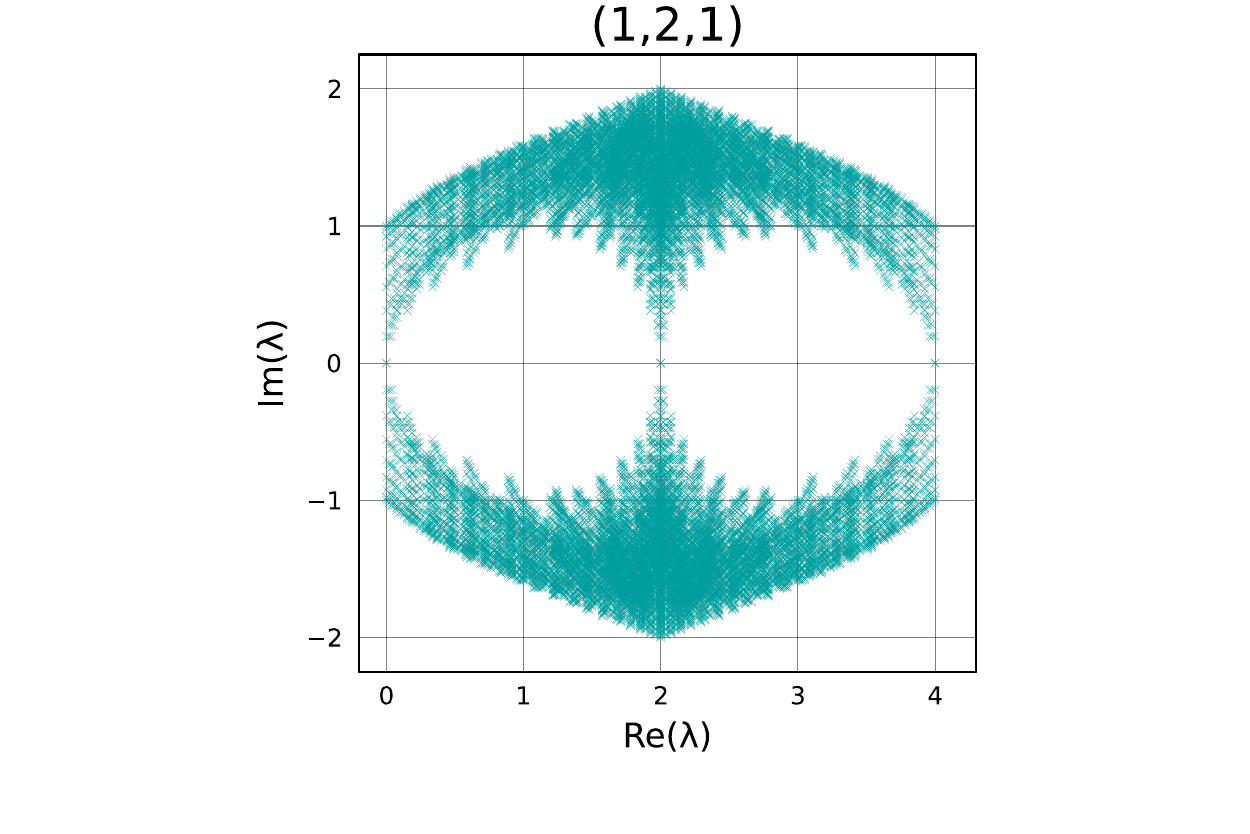}
                \subcaption*{$M_{12}+M_{13}$\\ 
                \vspace*{0.2cm} 
                $\{R^2_{13}C_0, R^2_{24}C_0,S_{\mu}I_S\}$}
            \end{subfigure}
            \hspace{-1cm}
            \begin{subfigure}{.5\textwidth}
                \centering
                \includegraphics[trim={3cm 0.8cm 2cm 0},clip,width=0.85\textwidth]{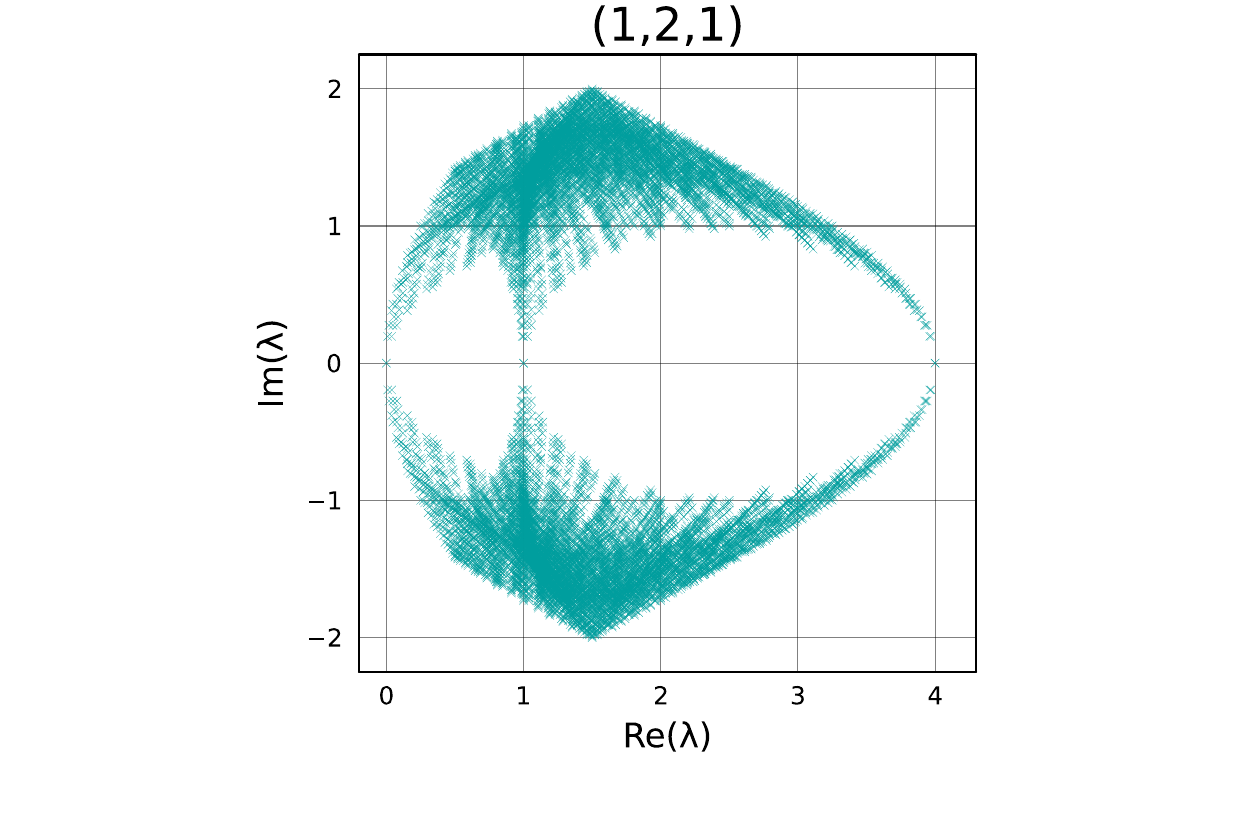}
                \subcaption*{$\frac{1}{2} M_{12}+M_{34}+M_{A}\sim\diag(-0.5,1.5,-0.5,2.5)$\\
                \vspace*{0.2cm} 
                $\{R^2_{13}C_0, R^2_{24}C_0,S_{\mu}I_S,R_{\mu\nu}R_{\rho\sigma}\}$}
            \end{subfigure}
            \caption{1-flavored mass terms and their symmetries by a 1+2+1 splitting.}
            \label{fig:1+2+1}
            \end{figure}

\begin{figure}[H]
            \captionsetup[subfigure]{justification=centering}
            \centering
            \begin{subfigure}{.48\textwidth}
                \centering
                \includegraphics[trim={5.5cm 0.8cm 4cm 0},clip,width=0.7\textwidth]{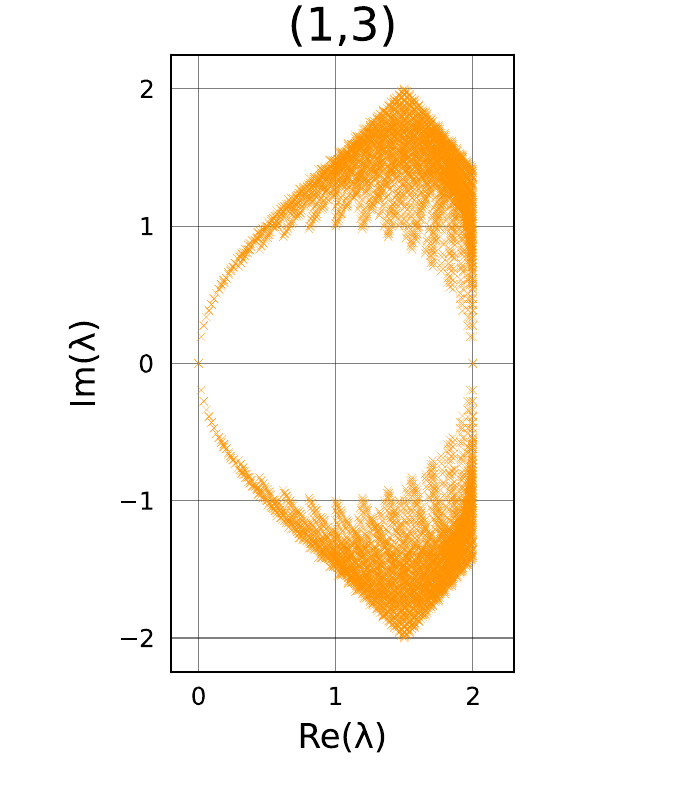}
                \subcaption*{$-M_{12}+M_{34}+M_A\sim\diag(1,-3,1,1)$\\
                \vspace*{0.2cm} $\{R^2_{13}C_0, R^2_{24}C_0,S_{\mu}I_S,R_{\mu\nu}R_{\rho\sigma}\}$}
            \end{subfigure}
            \hspace*{-1.8cm}
            \begin{subfigure}{.48\textwidth}
                \centering
                \includegraphics[trim={5.5cm 0.8cm 4cm 0},clip,width=0.7\textwidth]{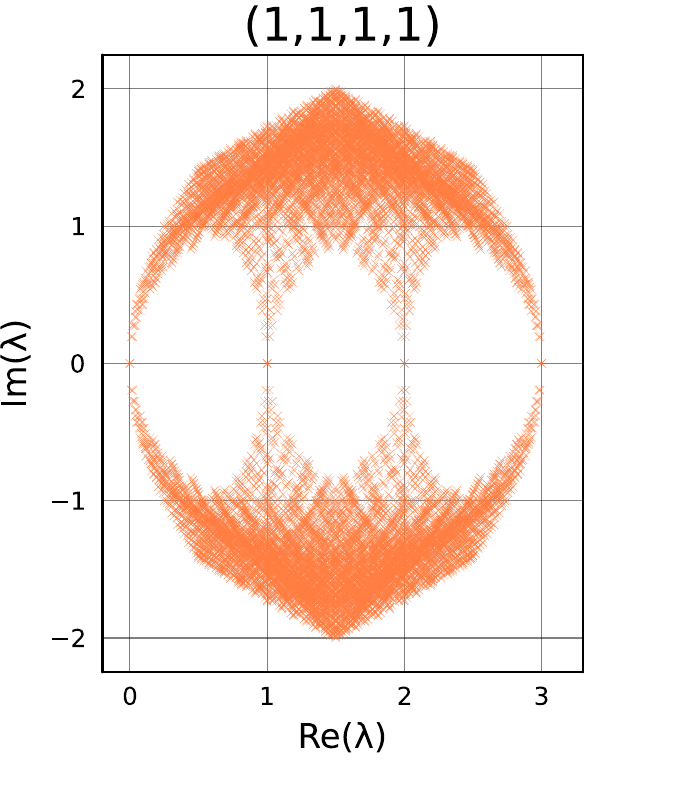}
                \subcaption*{$\frac{1}{2} M_{12}+M_{34}\sim\diag(0.5,-0.5,-1.5,1.5)$ \\ 
                \vspace*{0.2cm} $\{R^{2}_{13}C_0, R^{2}_{24}C_0, S_{\mu}I_S,R_{\mu\nu}R_{\rho\sigma}\}$}
            \end{subfigure}
            \caption{1-flavored mass terms and their symmetries from a 1+3 and 1+1+1+1 splitting.}
            \label{fig:others}
            \end{figure}

The staggered action is invariant under the following discrete symmetries \cite{Toolkit,LectureNotes}:
\begin{enumerate}
      \item
      Rotations  
      \vspace*{-0.9cm}
      \begin{align}
          R_{\mu\nu}: \chi(n) \to S_R\left(R^{-1}n\right) \chi\left(R^{-1}n\right)
      \end{align}
      \item 
      Shifts 
      \vspace*{-0.9cm}
       \begin{align}
          S_{\mu}: \chi(n) \to \zeta_{\mu}(n) \chi\left(n + \hat{\mu}\right)
      \end{align}
        \item 
      Spatial inversion
      \vspace*{-1.0cm}
       \begin{align}
          I_S: \chi(n) \to \eta_{4}(n) \chi\left(I_S^{-1}(n)\right)
      \end{align}
        \item Charge conjugation 
        \vspace*{-1.2cm}
       \begin{align}
          C_0: \begin{cases}
         \chi(n) \to \epsilon(n) \Bar{\chi}\left(n\right)\\
         \Bar{\chi}(n) \to -\epsilon(n) \chi\left(n\right)
          \end{cases}
      \end{align}
  \end{enumerate}
  where $\zeta_{\mu}=(-)^{\sum_{\nu > \mu}n_{\nu}}$. Under rotation $R_{\mu\nu}$, spacetime indices transform as $\mu \rightarrow \nu$, $\nu \rightarrow -\mu$, $\tau \rightarrow \tau$ for $\tau \neq \mu, \nu$ and the phase 
   \begin{align}
    S_R(n) = \frac{1}{2}\left(1 \pm \eta_{\mu}\eta_{\nu} \mp \zeta_{\mu}\zeta_{\nu} + \zeta_{\mu}\zeta_{\nu} \eta_{\mu}\eta_{\nu} \right)\quad (\mu \lessgtr \nu).
    \label{def:S_R}
\end{align}
   Introducing taste splitting masses to staggered fermions partially breaks these symmetries into subgroups as shown by Misumi et al. \cite{Misumi}. The following table summarizes these symmetries for the staggered operator, the Adams operator $M_{A}$,  and the single taste combination $M_{H}=M_{\mu\nu}+M_{\rho\sigma}$. 
   \begin{table}[H]
        \centering
        \begin{tabular}{ c|c|c|c }
        \hline
        type  & $R_{\mu\nu}$ subgroups & $S_{\mu} \& I_{S}$ subgroups & $C_0$ subgroups \\
        \hline \hline
        staggered  & $R_{\mu\nu}$ & $S_{\mu}$, $I_{S}$ & $C_0$\\
        Adams & $R_{\mu\nu}$ & $S_{\mu} S_{\nu}$, $S_{\mu} I_{S}$ & $C_0$\\
        single taste & $R_{\mu\nu}R_{\rho\sigma}$ & $S_{\mu} S_{\nu} S_{\rho} S_{\sigma}$, $S_{\mu} I_S$ & $R_{\mu\rho}^{2}C_{0}$\tablefootnote[3]{Note that this charge conjugation operator is for the $M_{H}=M_{\mu\nu}+M_{\rho\sigma}$ type mass term and therefore differs from the one given in \cite{Misumi}, where a single flavor mass term of the form
        $M_{H}=\frac{1}{\sqrt{3}}(M_{12}+M_{34}+M_{13}+M_{42}+M_{14}+M_{23})$
    was considered.}\\
        \hline
        \end{tabular}
            \caption{Symmetries of staggered, Adams, and single taste type operators.}
    \end{table}
More general combinations of mass terms are of course possible. To provide a brief overview, we plot the free eigenvalue (EV) spectra for various combinations of mass terms in \autoref{fig:2+2} to \ref{fig:others}.

        \vspace*{-0.1cm}
      
The 2 + 2 splitting shown in \autoref{fig:2+2} gives a clear picture of how the splitting is only partial for a single mass term. A single 2-hop term splits the tastes into 2+2 while leaving the momentum in two directions unmodified, resulting in the less degenerate EVs of the spectrum along the imaginary axis as seen for $M_{12}$ compared to $M_{A}$. Moving on to the 1+2+1 splitting in \autoref{fig:1+2+1}, the different symmetry breaking can be seen in particular comparing the $M_{12}+M_{13}$ and $M_{A}+M_{12}$ spectra, where the EVs at the outer most branches of $M_{12}+M_{13}$ are less degenerate. The spectra show more spikes rather than smooth rounded inner circles compared to $M_{12}+M_{34}$. The spectrum of $\frac{1}{2}M_{12}+M_{34}+M_{A}$ amplifies the non-degeneracy of EVs and sharp cuts and brings an additional splitting of non-equidistant branches. A single taste operator can also be achieved by a 1+3 resp. 1+1+1+1 splitting as shown in \autoref{fig:others}. The EVs are once more less degenerate at the non-physical branch of the 1+3 operator,  while the 1+1+1+1 operator exhibits sharp edges appearing due to the larger prefactor $M_{34}$. Overall, it is evident that less symmetry leads to a smaller degeneracy in the free eigenvalue spectra. \\
A peculiar consequence of the rotational symmetry breaking in particular is the appearance of gluonic counterterms, as pointed out by Sharpe \cite{Sharpe}. Different coefficients appear in the gluonic action depending on the mass term taken. In our case, for 
\begin{align}
    M_{H}=M_{12}+M_{34}
    \label{eq:one-flavor-mass}
\end{align}
we need to separate between
\begin{align}
    F^{2}_{12}+F^{2}_{34} \quad \text{and} \quad F^{2}_{13}+F^{2}_{24}+F^{2}_{14}+F^{2}_{23}.
\end{align} 
More generally for $M_{\mu\nu}+M_{\alpha\beta}$ with all indices different, the counterterms are of the form
\begin{align}
        a(F_{\mu\nu}^2+F_{\alpha\beta}^2)+b(F_{\mu\alpha}^2+F_{\nu\beta}^2+F_{\mu\beta}^2+F_{\nu\alpha}^2).
        \label{eq:counterterms}
    \end{align}
Note that this symmetry breaking effect enters via the fermion determinant and thus is relevant only when unquenching. 

\section{Numerical results}
We present first numerical results on quenched configurations. The eigenvalue spectra give insight into the severity of the rotational symmetry breaking for practical purposes. Starting off with pure gauge configurations, which were kindly provided by Timo Eichhorn \cite{Timo,Timoetal}, the numerical setup is listed below. The configuration where generated for different $\beta$, up to 35  stout smearing steps with $\rho=0.12$ with an update algorithm of one heat bath followed by four overrelaxation sweeps, the scale is taken from \cite{Durretal}. The simulation parameters are as follows:

    \begin{itemize}[noitemsep]
        \item[$\circ$] $4^4, 6^4, 8^4$ lattices
        \item[$\circ$] $\beta\in \{5.7,5.8,5.9,6.0\}$
        \item[$\circ$] $a(\SI{}{\femto\meter})\in \{0.168, 0.133, 0.109, 0.091\}$
        \item[$\circ$] Up to 35 stout smearing steps, $\rho=0.12$
        \item[$\circ$] 20 configurations.
    \end{itemize}

We compare spectra of staggered operators with a mass term given in \autoref{eq:one-flavor-mass} to those with a mass term rotated by a $R_{23}$ rotation
\begin{align}
        M_{12}+M_{34} \xrightarrow{R_{23}} M_{13}+M_{42}.
    \end{align} 

    \begin{figure}[H]
            \centering
            \begin{subfigure}{.5\textwidth}
                \centering
                \includegraphics[width=0.9\textwidth]{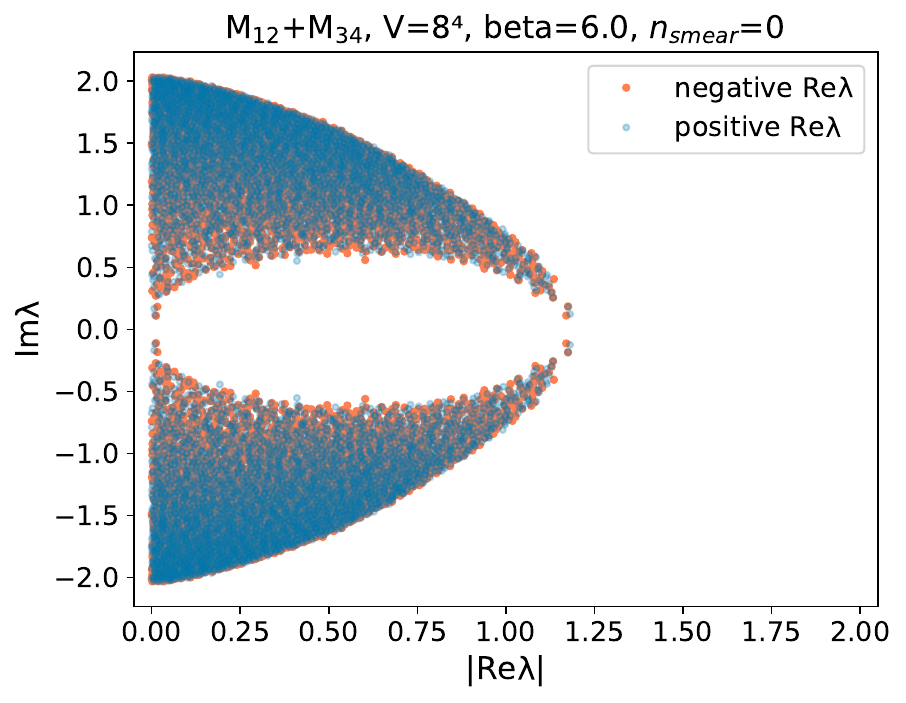}
                \subcaption*{}
            \end{subfigure}
            \hspace{-0.2cm}
            \begin{subfigure}{.5\textwidth}
                \centering
                \includegraphics[width=0.9\textwidth]{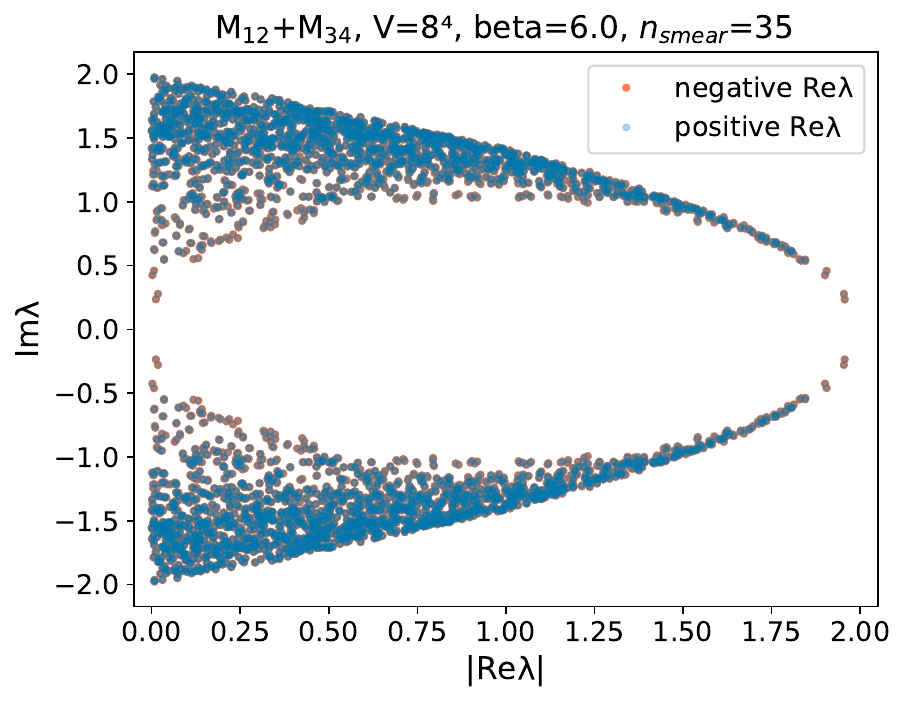}
                \subcaption*{}
            \end{subfigure}
            \vspace*{-0.5cm}
            \caption{Folded eigenvalue spectra for $V=8^4$ at $\beta=6.0$ and $n_{\text{smear}}=0$ and $n_{\text{smear}}=35$.}
            \label{fig:EVspec}
    \end{figure}
\begin{wrapfigure}[8]{l}{0.5\textwidth}
        \centering
        \includegraphics[width=0.45\textwidth]{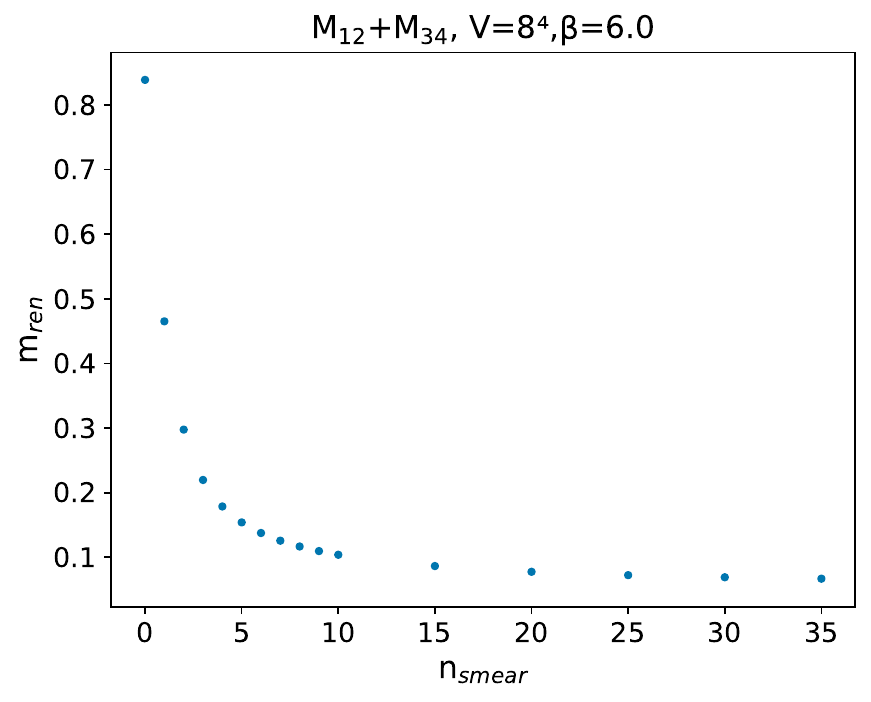}
        \caption{Additive mass renormalization, determined by the distance of the spectrum from the real axis, versus the number of smearing steps.}
        \label{fig:MassRen}
    \end{wrapfigure}

   In \autoref{fig:EVspec}, we show the change in the EV spectrum of $M_{12}+M_{34}$ with increasing smearing steps. The spectra are folded showing the overlap of eigenvalues with positive and negative real part. Symmetry breaking is evident for unsmeared spectra where the positive and negative real parts of the eigenvalues don't match. When smearing the gauge configurations, the eigenvalues stretch out further across the real axis and the symmetry breaking is milder. We also look at the distance of the smallest eigenvalues from the real axis, which is a simple proxy for the additive mass renormalization. In  \autoref{fig:MassRen} we plot this number averaged over 20 configurations versus the number of smearing steps. The standard error bars are too small to be displayed. 

\begin{figure}[H]
            \captionsetup[subfigure]{justification=centering}
            \centering
            \begin{subfigure}{.5\textwidth}
                \centering
                \includegraphics[width=0.9\textwidth]{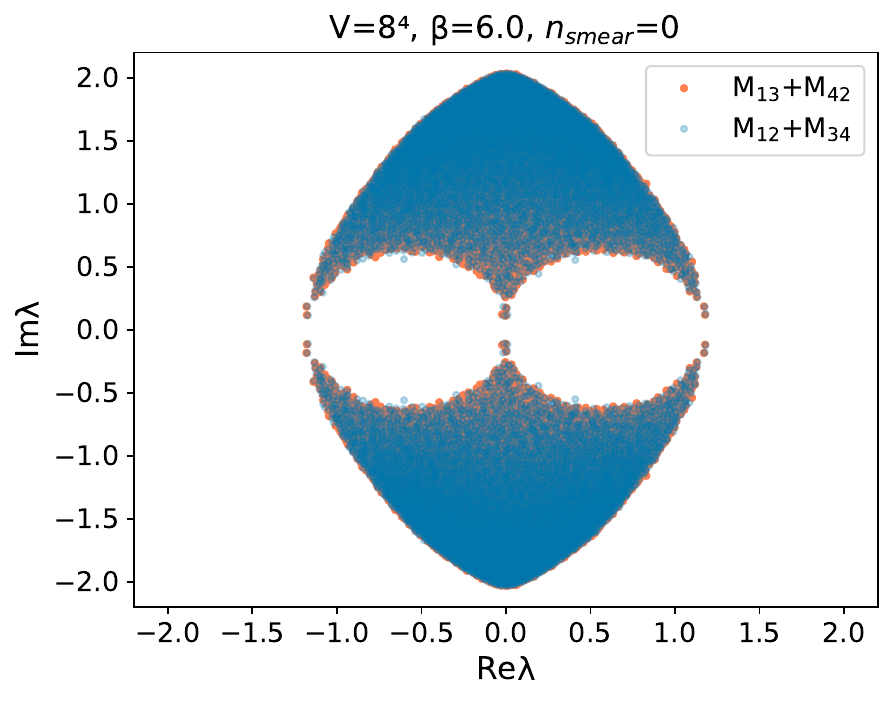}
                \subcaption*{}
            \end{subfigure}%
            \begin{subfigure}{.5\textwidth}
                \centering
                \includegraphics[width=0.9\textwidth]{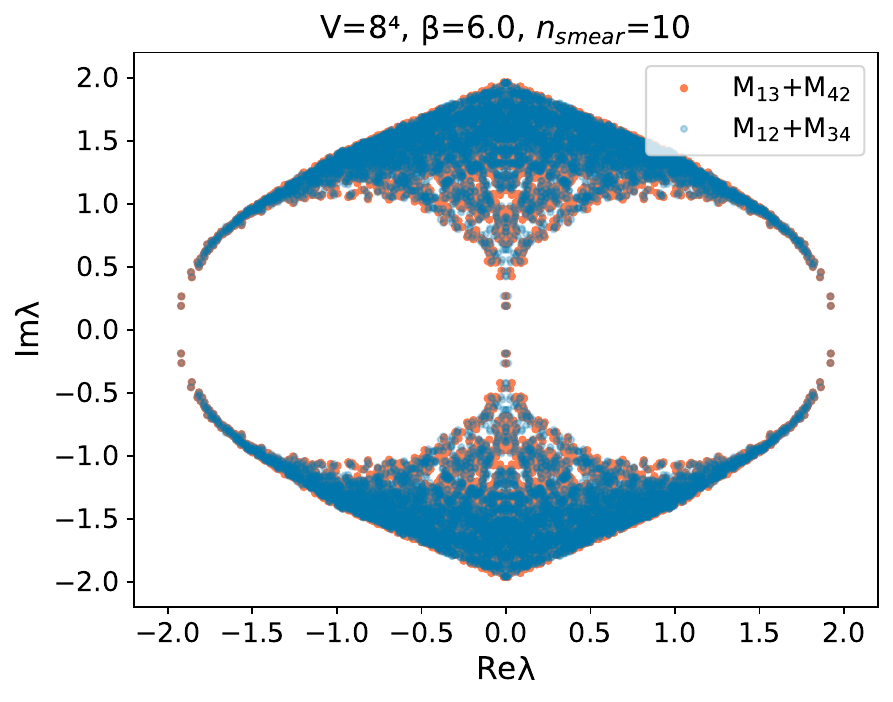}
                \subcaption*{}
            \end{subfigure}%
             \vspace*{-0.5cm}
            \caption{Eigenvalue spectra of $M_{12}+M_{34}$ and $M_{13}+M_{42}$ for $V=8^4,\beta=6.0$ and smearing steps $n_{\text{smear}}=0$ and $n_{\text{smear}}=10$.}
            \label{fig:compare}
    \end{figure}

    \vspace*{0.3cm}
Considering the full rotation of the mass term, the observed spectra shown in \autoref{fig:compare} of $M_{12}+M_{34}$ and $M_{13}+M_{42}$ indicate substantially milder rotational symmetry breaking after smearing, especially in the physical branch and for near-zero modes, consistent with the expectations from the eigenvalue spectrum \autoref{fig:EVspec}. 

\newpage

    \begin{wrapfigure}[15]{l}{0.48\textwidth} 
        \centering
        \includegraphics[width=0.45\textwidth]{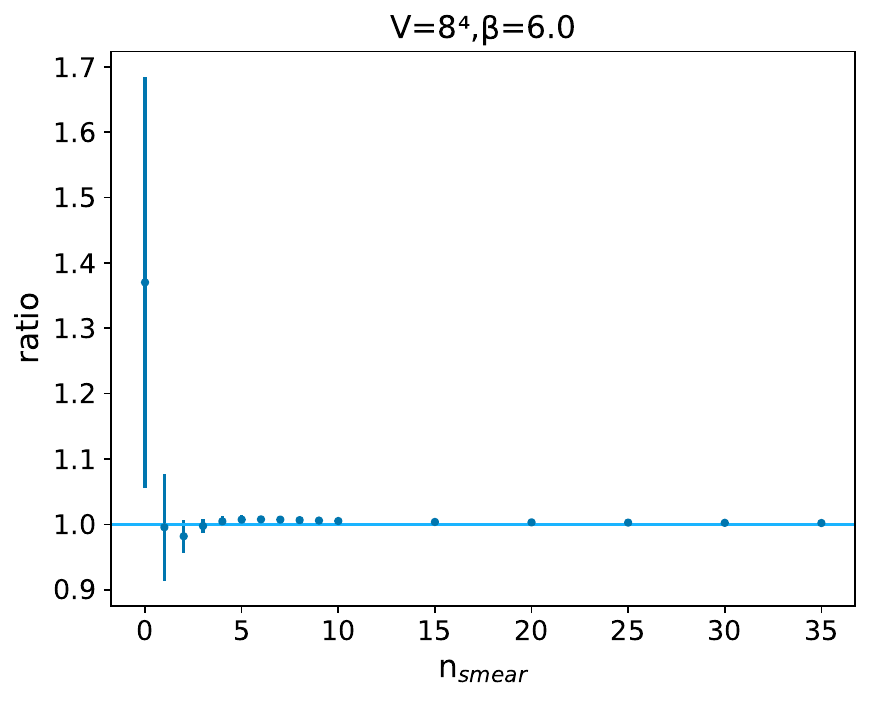}
        \caption{Determinant ratio for $V=8^4$at $\beta=6.0$ with respect to the number of smearing steps.}
        \label{fig:determinant}
    \end{wrapfigure}    
        
\begin{figure}[H]
        \vspace*{-0.5cm}
            \centering
                \centering
                \includegraphics[width=0.5\textwidth]{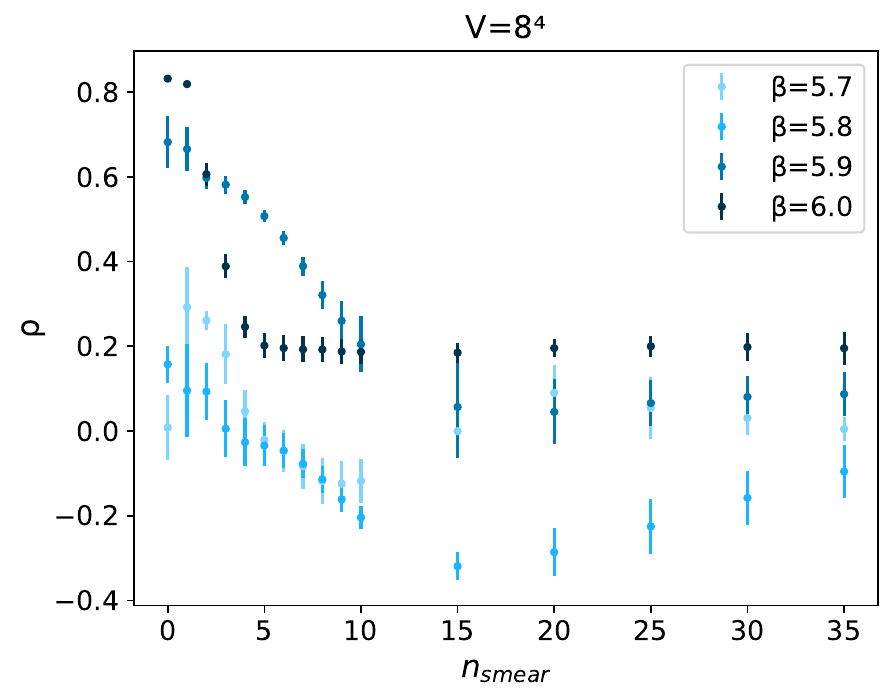}
            \vspace*{-0.3cm}
            \caption{Correlation coefficient for $V=8^4$ and different $\beta$ for an increasing number of smearing steps.}
            \label{fig:Counterterms}
        \end{figure}

In order to quantify this symmetry breaking we compute the ratio of the determinants   
    \begin{align}
        \text{ratio} = \frac{\det(D_{st}+M_{12}+M_{34})}{\det(D_{st}+M_{13}+M_{42})}
        \label{eq:ratio}
    \end{align}
with $D_{\text{stag}}$ the staggered operator. \autoref{fig:determinant} shows that the average determinant ratio is compatible with 1 for all smearing levels, indicating that the effects of rotational symmetry breaking are not very pronounced, at least for the very small volumes we studied. The standard errors decrease exponentially with higher smearing steps. \newline
\newline 
Lastly, we consider the correlation of the determiant ratio to the gluonic counterterms emerging for the considered mass terms. From \autoref{eq:counterterms} we find the counterterm structure dependency to be of the form
    \begin{align}
        \frac{\det(D_{st}+M_{12}+M_{34})}{\det(D_{st}+M_{13}+M_{42})} \sim (a-b)(F_{12}^2+F_{34}^2)-(a-b)(F_{13}^2+F_{24}^2)
        \label{eq:counterterms_U_intermediate}
    \end{align}
    and ultimately expressed in terms of the plaquette $U_{\mu\nu}$ 
    \begin{align}
        \frac{\det(D_{st}+M_{12}+M_{34})}{\det(D_{st}+M_{13}+M_{42})} \sim (U_{12}+U_{34})-(U_{13}+U_{24}).
        \label{eq:counterterms_U}
    \end{align}
    \autoref{fig:Counterterms} shows the correlation coefficient $\rho$ of the ratio given in \autoref{eq:counterterms_U} for different $\beta$ versus the number of smearing steps. 
\clearpage
    
The right hand side of \autoref{eq:counterterms_U} is computed for unsmeared configurations. The correlation between the determinant ratios and the corresponding counterterm structure confirms the dependence for unsmeared configurations. The more the gauge configurations are smeared, the smaller the correlation. After 10 smearing steps, the correlation coefficient predominantly remains constant up to $n_{\text{smear}}=35$. The errorbars on the first two data points of $\beta=6.0$ are too small to be displayed.

\section{Conclusion and Outlook}
Given the symmetry properties of taste splitting masses for staggered fermions as summarized in \cite{Misumi}, we have studied these properties numerically in the free case as well as for pure gauge configurations. In particular, the rotational symmetry was studied in order to understand the severity of the breaking for single taste masses and to gain insights into the structure of counterterms emerging from it. First studies of the eigenvalue spectra show that rotational symmetry breaking is milder for stout smeared configurations while the additive mass renormalization is smaller. Furthermore, the determinant ratio of the single taste operator $M_{12}+M_{34}$ and the operator resulting from rotation $M_{13}+M_{42}$ not only confirm the improved symmetry after smearing, but show that even without smearing the breaking of rotational symmetry is not too severe. Finally, the correlation coefficient of the determinant ratio to the general counterterm structure indicates, that there is little correlation for highly smeared gauge configurations. 

These results give first insights into the severity of rotational symmetry breaking for single taste staggered fermion operators. Although our exploratory study was performed on quenched configurations in small volumes, it seems that at higher smearing levels the rotational symmetry breaking effects are manageable. As a next step, we plan to expand our investigation to unquenched gauge configurations, which will be generated in collaboration with Gianluca Fuwa \cite{Gianluca}.

\section*{Acknowledgement}
We thank Timo Eichhorn for providing the gauge configurations and  Stephan Durr, Fabian Frech as well as Lukas Varnhorst for insightful discussions. Nuha Chreim is supported by the Hans-Böckler-Stiftung under grant No. 415369.

\newpage
\fontsize{11}{12}\selectfont

\end{document}